\documentclass[runningheads]{llncs}
\usepackage{graphicx, amssymb,color,cite,array, url}
\usepackage{tikz}
\usepackage{pgfplots}

\begin{document}
\title{DeScarGAN: Disease-Specific Anomaly Detection with Weak Supervision }
\titlerunning{DeScarGAN}

\author{
Julia Wolleb \and
Robin Sandk\"uhler \and
Philippe C. Cattin}

\authorrunning{J. Wolleb et al.}

\institute{Department of Biomedical Engineering, University of Basel, Allschwil, Switzerland\\
\email{julia.wolleb@unibas.ch}}

\maketitle             

\begin{abstract}
Anomaly detection and localization in medical images is a challenging task, especially when the anomaly exhibits a change of existing structures, e.g., brain atrophy or changes in the pleural space due to pleural effusions. In this work, we present a  weakly supervised and detail-preserving method that is able to detect structural changes of existing anatomical structures. In contrast to standard anomaly detection methods, our method extracts information about the disease characteristics from two groups: a group of patients affected by the same disease and a healthy control group. Together with identity-preserving mechanisms, this enables our method to extract highly disease-specific characteristics for a more detailed detection of structural changes. We designed a specific synthetic data set to evaluate and compare our method against state-of-the-art anomaly detection methods. Finally, we show the performance of our method on chest X-ray images. Our method called DeScarGAN outperforms other anomaly detection methods on the synthetic data set and by visual inspection on the chest X-ray image data set.
\end{abstract}

\section{Introduction}
For medical applications, it is of great interest to find an automated way to show visual manifestations of a disease. In the past, artificial neural networks have shown a great performance in the task of image segmentation. As the manual generation of pixel-wise annotations is time consuming and requires expert knowledge, the training data is limited in number or even unavailable. Furthermore, the manually generated labels are affected by human bias. Using only image-level class labels for training of the networks overcomes those issues.
In this paper, we propose a new disease-specific and weakly supervised method for anomaly detection and localization. The task we aim to solve is to highlight the pathological changes in an image of a diseased subject, as well as the classification into diseased and healthy subjects. This can improve diagnosis, lead the attention to relevant parts of the anatomy and provide a starting point for further studies.

Classical anomaly detection algorithms are trained only on healthy subjects and detect abnormal parts of images as outliers. Variational Autoencoders (VAEs) can be used to detect lesions in the brain \cite{chen, zimmerer}. Beside VAEs, Generative Adversarial Networks  (GANs) \cite{gan} are used for anomaly detection in medical images \cite{survey}. %, anogan, ganomaly, skip-ganomaly}. 
VAGAN \cite{vagan} proposes the generation of an additive map to make an image of a diseased subject appear healthy.
PathoGAN \cite{pathogan} provides a weakly supervised segmentation algorithm for brain tumors based on image-to-image translation. StarGAN \cite{star} follows a similar idea as CycleGAN \cite{ciclegan} and simplifies the architecture to only one generator and one discriminator. This idea can be used for anomaly detection by taking the difference between original and translated images. Fixed-Point GAN (FP-GAN) \cite{fixedpoint} improves StarGAN by preserving features that should not be changed during translation, outperforming f-Anogan \cite{f-ano} and others \cite{gan-lesion} in brain lesion detection.
The problem of combining GANs with a classification network is tackled by semi-supervised GANs \cite{Odena, Saliman}. Class activation maps \cite{cam, gradcam} visualize the features of the input image that lead to the classification score, but limitations in the resolution lead to blurry maps.  Another approach is the generation of saliency maps \cite{saliency, saliency_seg} by computing the gradient of the classification score with regard to the input image.\\
We are interested in cases where the anomaly occurs in the form of deformations of existing structures, e.g., atrophy, rather than in lesions. Both VAGAN and FP-GAN are designed to only generate an additive map rather than a complete new image. We claim that this restriction to additive maps may hinder the methods from showing deformations. What is more, VAGAN is not designed to perform classification and assumes that the class label for each input image is provided in advance. VAEs are only trained on the healthy control group and may not be able to point out the characteristics of a specific disease, due to natural variations in the data. Our method for detection of structural changes in anatomical regions, further called DeScarGAN, is designed to address these issues.

Our method performs image-to-image translation between a set of healthy and a set of diseased subjects in order to find the visual manifestations that make the distributions of the two datasets differ from each other. We introduce a novel disease-specific architecture with skip connections, a splitting of the networks into weight-sharing subnetworks and an identity loss as identity-preserving mechanisms. This ensures that the difference between the generated healthy and the real input image is accurate enough to highlight the regions of interest, resulting in more detailed maps of the characteristics of the disease than previous methods.\\
We point out that compared to classical anomaly detection, we train only on one specific disease and extract information about its characteristics.  With this approach, changes of already existing structures can be detected in a detailed manner, which is different from the presence or absence of lesions. \\
We evaluate our method on a synthetic dataset designed for this task. Furthermore, we apply it on the Chexpert dataset \cite{chex} of X-ray images of lungs in order to detect pleural effusions. Our method outperforms state-of-the-art anomaly detection algorithms in showing deformations of already existing structures. Furthermore,  it provides better classification results than standard classification algorithms. With the addition of visually highlighting the regions of interest, the attention is led to the relevant parts of the image, making a step towards \emph{interpretable machine learning}. The code is publicly available at \url{https://github.com/JuliaWolleb/DeScarGAN}.

\section{Method}

\begin{figure}
 	\centering
 	\includegraphics[width=0.88\textwidth]{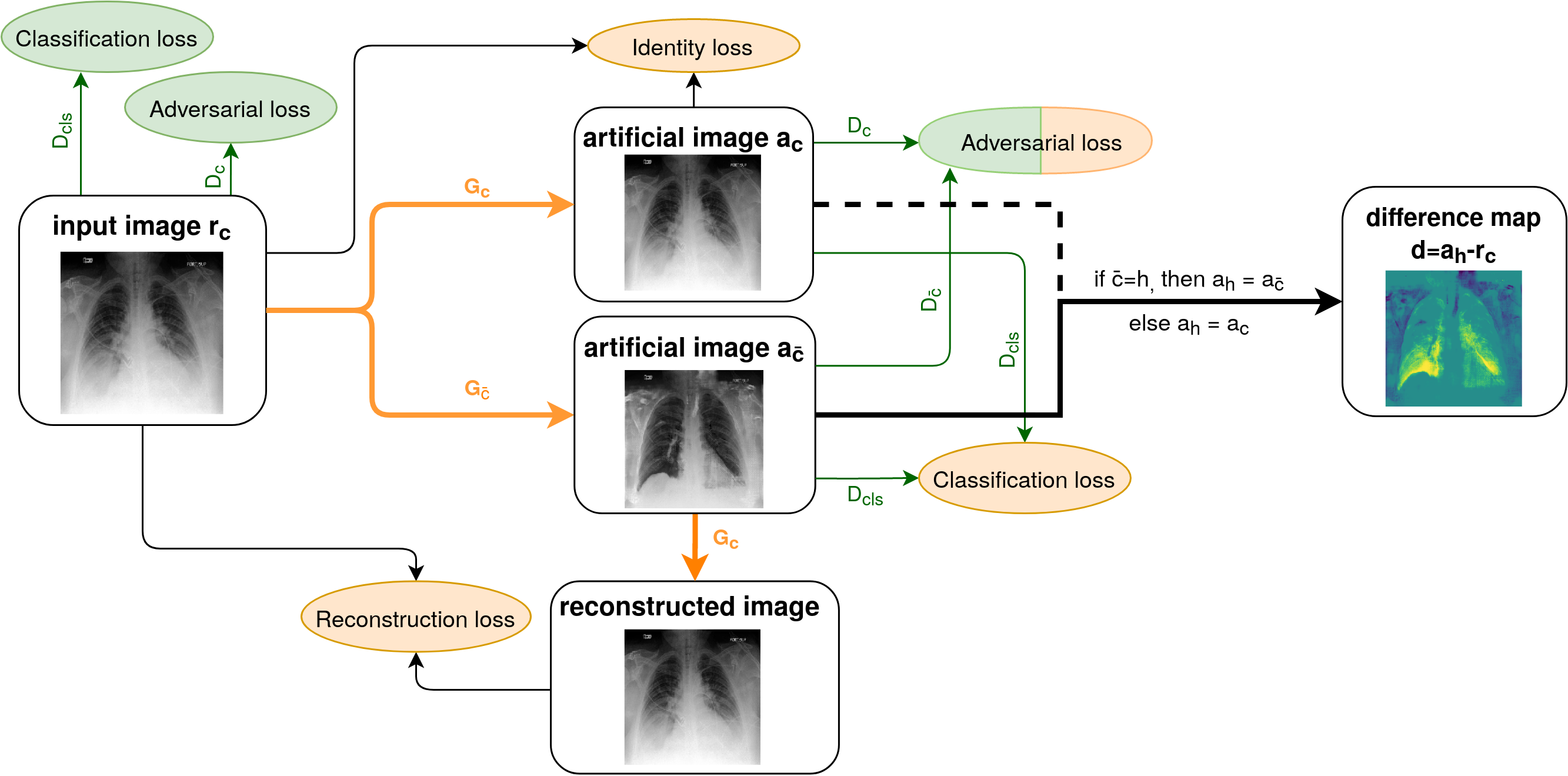}
 	\caption{Workflow of our method. The components of the loss functions for the discriminator are shown in green, the ones for the generator in orange.} \label{fig1}
 \end{figure}

Let  ${\mathcal{F} = \{x \; | \; x:  \mathbb{R}^2 \rightarrow  \mathbb{R} \}}$ be a set of medical images from the same imaging modality showing the same anatomical structures, with $\mathcal{P} \subset \mathcal{F}$ the set of images of patients affected by a specific disease and ${\mathcal{H}\subset \mathcal{F}}$  the set of images of a healthy control group. The aim of our method is, given a new image of unknown class, to detect regions in the image that show the same characteristics as the images in $\mathcal{P}$ and to assign a class label. \\
 Let $p$ be the class of the images in $\mathcal{P}$ and $h$ the class of the images in $\mathcal{H}$, with ${c, \bar{c} \in \{h, p\}}$ and ${c \neq \bar{c}}$. The main idea is to translate a real image $r_c$ of either class $c$ to an artificial image $a_{\bar{c}}$ of class $\bar{c}$. The pathological region is then defined as the difference ${d:=a_h-r_c}$ between the artificial healthy image $a_h$ and the real input image $r_c$ of class $c$.  
 Thus we perform image-to-image translation between the unpaired sets $\mathcal{P}$ and $\mathcal{H}$. A diagram showing the workflow of our method is given in Figure~\ref{fig1}.  Given any image $r_c$, the generator both generates an artificial image $a_c$ of the same class $c$ and an artificial image $a_{\bar{c}}$ of class $\bar{c}$. To ensure that $r_c$ and $a_h$ only differ in the pathological region, we add the identity loss $\mathcal{L}_{id}$ and the reconstruction loss $\mathcal{L}_{rec}$ for cycle consistency. 
 
\begin{figure}
 	\centering
 	\includegraphics[width=0.92\textwidth]{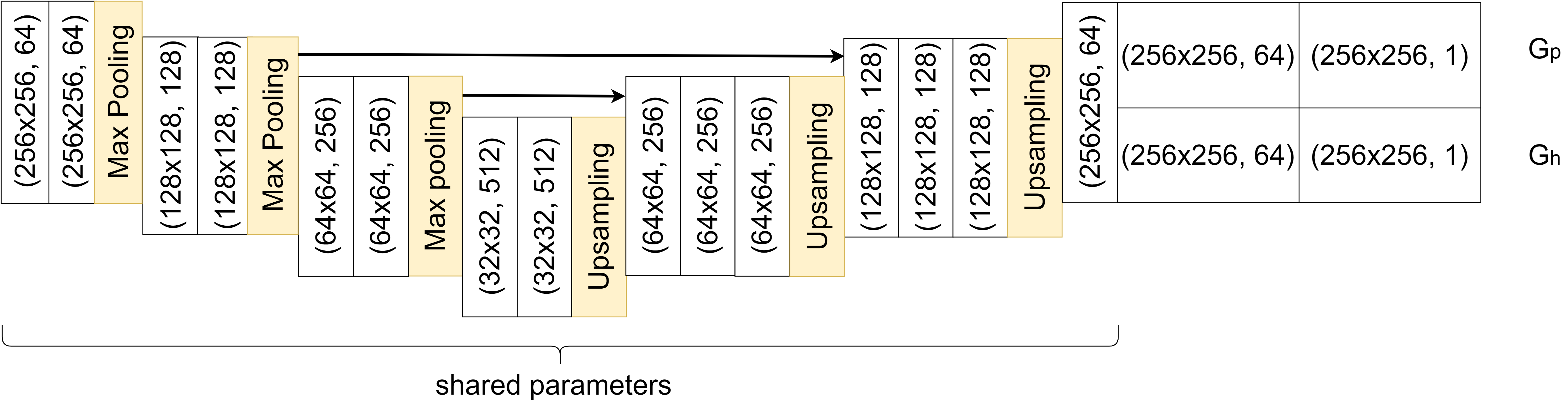}
 	\caption{The architecture of the generator network. Every box stands for a convolutional layer with the stated output size (image width $\times$ image height, feature channels) and kernelsize 3, followed by a batch normalization layer and a ReLU activation function.} \label{fig6}
 \end{figure}

The generator consists of two branches, its architecture is shown in Figure~\ref{fig6}. We refer to the generator ${G_p:\mathcal{F} \rightarrow \mathcal{P}} $ for generating images of class $p$ and generator ${G_h:\mathcal{F} \rightarrow \mathcal{H}}$ for generating images of class $h$. 

 The skip connections of the generator ensure that the artificial image maintains the detailed structures of the input image. This is a way to alter only the necessary features, thus making the difference map $d$ more accurate. The skip connection in the uppermost layer turned out to be too restrictive to perform the translation to another class. By omitting this skip connection, we enable the generator to perform structural changes. 
 
\begin{figure}
	\centering
	\includegraphics[width=0.92\textwidth]{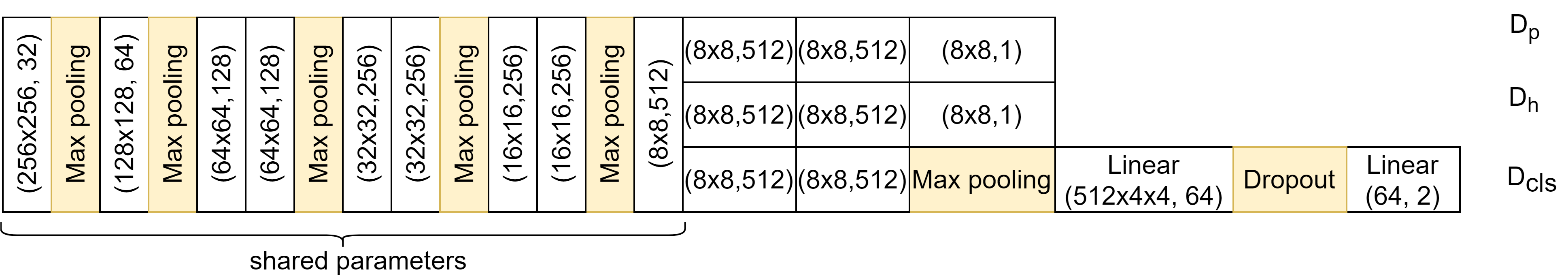}
	\caption{The architecture of the discriminator with three subnets $D_p$, $D_h$ and $D_{cls}$ that share parameters. Every box stands for a convolutional layer with kernelsize 3 and with the stated output size, followed by a ReLU activation function.} \label{fig5}
\end{figure}

The discriminator network has the task to both classify images into healthy and diseased subjects and to distinguish between real and artificial images. Therefore, it consists of three subnets that share parameters, as shown in Figure~\ref{fig5}. ${D_p:\mathcal{P} \rightarrow \mathbb{R}}$ distinguishes between real and artificial images of class $p$,  ${D_{h}:\mathcal{H} \rightarrow \mathbb{R}}$ does the same for class $h$ and ${D_{cls}:\mathcal{F} \rightarrow \mathbb{R}}$ is the network for classification, following the structure of a VGG net \cite{vgg}. The branching of the generator and discriminator gives a higher range of flexibility compared to StarGAN, which turned out to be beneficial for image-to-image translation.

With the notation from above,  $D_{c}$ can be  $D_{p}$ or $D_{h}$ interchangeably, and $D_{\bar{c}}$ denotes the discriminator for the contrary class. The same applies for the generator $G$. 

\subsection{Loss functions}

\subsubsection*{Adversarial Loss}
The generator aims to generate images that the discriminator cannot distinguish from real images. Following the idea of Wasserstein GANs   \cite{wgan}, we add a gradient penalty loss and define the adversarial loss for the discriminator as 
\begin{equation}
\mathcal{L}_{adv,d} = -\mathbb{E}_{r_c,c}[(D_{c}(r_c)) ] +\mathbb{E}_{r_c,\bar{c}}[D_{\bar{c}}(G_{\bar{c}}(r_c) ]+ \lambda_{gp} \mathbb{E}_{\hat{x},c}[(\parallel \nabla_{\hat{x}} D_c(\hat{x}_c) \parallel_{2}-1)^{2} ] , %identity loss!
\end{equation}
where $\hat{x}_c$ is given by $\hat{x}_c=tr_c+(1-t)a_{c}$  with $ t \thicksim U([0,1]).$
The adversarial loss for the generator is defined as 
\begin{equation}
\mathcal{L}_{adv,g} =- \mathbb{E}_{r_c,{\bar{c}}}[D_{\bar{c}}(G_{\bar{c}}(r_c) ].
\end{equation}

\subsubsection*{Identity Loss}
Considering an input image $r_c$, we aim for identity between $r_c$ and $G_c(r_c)$. Therefore, the identity loss for the generator is defined as

\begin{equation}
\mathcal{L}_{id} = \mathbb{E}_{r_c,c}[\parallel r_c-G_c(r_c) \parallel_{2} ]. %identity loss!
\end{equation}

\subsubsection*{Classification Loss}
 The classification subnet $D_{cls}$ of the discriminator has to correctly classify $r_c$ to belong to class $c$. The objective function for the discriminator is described as
\begin{equation}
\mathcal{L}_{cls,d} = \mathbb{E}_{r_c,c}[- \textrm{log}D_{cls}^c(r_c) ],
\end{equation}
where the term $D_{cls}^c(r_c)$ describes the computed probability score that $r_c$ belongs to class $c$. 
The generator aims for classification of an artificial image $a_{\bar{c}}=G_{\bar{c}}(r_c)$ to belong to class $\bar{c}$. Therefore, the classification loss for the generator is defined as
\begin{equation}
\mathcal{L}_{cls,g} = \mathbb{E}_{r_c,\bar{c}}[- \textrm{log} D_{cls}^{\bar{c}}( G_{\bar{c}}(r_c)) ].
\end{equation}
\subsubsection*{Reconstruction Loss}
When an input image $r_c$ of class $c$ is translated into an image $a_{\bar{c}}=G_{\bar{c}}(r_c)$ of class ${\bar{c}}$, we aim for cycle consistency when translating $a_{\bar{c}}$ back to class $c$. This is achieved by adding a reconstruction loss term for the generator, given by

\begin{equation}
\mathcal{L}_{rec} = \mathbb{E}_{r_c,c}[\parallel r_c-G_c(G_{\bar{c}}(r_c)) \parallel_{2} ] .
\end{equation}
\subsubsection*{Total Loss Objective}
The overall loss function for the generator is defined as

\begin{equation}
\mathcal{L}_{g}=\lambda_{adv,g}\mathcal{L}_{adv,g}+\lambda_{rec}\mathcal{L}_{rec}+\lambda_{id}\mathcal{L}_{id}+\lambda_{cls,g}\mathcal{L}_{cls,g},
\end{equation}
and for the discriminator as
\begin{equation}
\mathcal{L}_{d}=\lambda_{adv,d}\mathcal{L}_{adv,d}+\lambda_{cls,d}\mathcal{L}_{cls,d}.
\end{equation}

\section{Synthetic Dataset}
The purpose of weakly supervised algorithms is to overcome the need for pixel-wise labels and the human bias within these labels. In order not to be affected by this human bias, we designed  a synthetic data set for the evaluation of our method. 
 Two ellipses $e_1$ and $e_2$ are present in the image, one larger than the other and both with variable contour thickness, origin and orientation. The background is structured in concentric waves with two variable origins and variable wave length; this provides a higher level of complexity. Images of the healthy group $\mathcal{H}$ keep this structure. If the image is deformed such that the smaller ellipse $e_1$ shrinks to an even smaller ellipse, the background is also deformed. Images with this characteristics belong to the diseased group $\mathcal{P}$. Implementation details are provided in the supplementary material.\\
  In Figure~$\ref{fig2}$, exemplary images of the two sets $\mathcal{H}$ and $\mathcal{P}$ are shown. The pixel-wise ground truth ($GT$) is known by definition.
We generate a training set of 2000 images of each class, and a validation and a test set with 200 images of each class. 

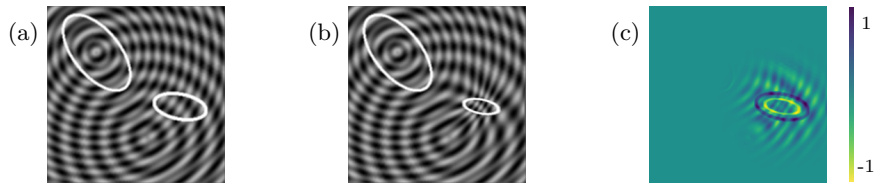
\begin{figure}

	\centering
	\begin{tikzpicture}
	\node[draw=white, inner sep=0pt, thick] at (1.5, 0) {\includegraphics[scale=0.26]{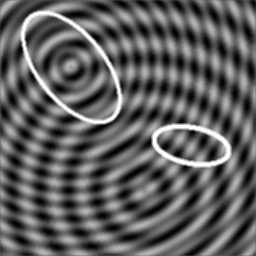}};
	\node[] at (0, 0.8) {\small(a)};
	\node[draw=white, inner sep=0pt, thick] at (5.5, 0) {\includegraphics[scale=0.26]{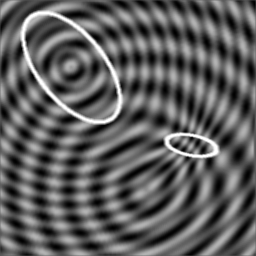}};
	\node[] at (4, 0.8) {\small(b)};
	\node[draw=white, inner sep=0pt, thick] at (9.5, 0) {\includegraphics[scale=0.26]{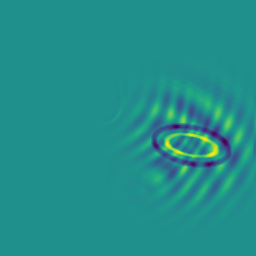}};
	\node[] at (8, 0.8) {\small(c)};
	\node[draw=white, inner sep=0pt, thick] at (11, 0) {\includegraphics[scale=0.071]{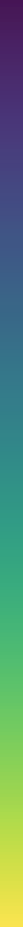}};
	\node[] at (11.2, 0.95) {\scriptsize 1};
	\node[] at (11.2, -0.95) {\scriptsize -1};
	\end{tikzpicture}

	\caption{Images (a) and (b) show exemplary images of the sets $\mathcal{H}$ and $\mathcal{P}$ respectively. Image (c) corresponds to the ground truth given by the difference  (a) - (b).}
	\label{fig2}
\end{figure}

\section{Results and Discussion}

We compare our method against  StarGAN, FP-GAN, the VAE proposed in \cite{chen} and VAGAN. To train our model, we use the Adam optimizer \cite{adam} with ${\beta_1=0.5}$, ${\beta_2=0.999}$, and a learning rate of ${10^{-4}}$. For every update of the parameters of the generator, we update the discriminator 5 times.  We manually choose the hyperparameters ${\lambda_{adv,d}=20}$, ${\lambda_{gp}=10, \, \lambda_{id}=\lambda_{rec}=50}$, ${\lambda_{adv,g}=\lambda_{cls,g}=1}$, and ${\lambda_{cls,d}=5}$. The number of trained parameters is 8528262 for the generator and 18170180 for the discriminator.

\subsection{Synthetic Dataset}

As a measure for the pixel-wise error for the anomaly detection task, we choose the Dice score,  AUROC$_{pix}$ \cite{auroc} for pixel-wise classification, the Mean Square Error (MSE) and the Structural Similarity Index (SSIM) between  $d$ and $GT$, and finally the  MSE between an input image $r_h \in \mathcal{H}$ and the corresponding artificial image $a_h$. For the calculation of the Dice score and the AUROC$_{pix}$, we perform a thresholding based on the average  Otsu \cite{ostu} threshold value on the $GT$ images. The results are shown in Table  $\ref{tab1}$. All methods classify the images almost perfectly on the test set, so we omit those results. VAGAN is not designed to take an image $r_h \in \mathcal{H}$ as input, but we still report the result for completeness. 
\begin{table}
	\caption{Results on the synthetic dataset.}\label{tab1}
	 \centering
\begin{tabular}{p{1.8cm} >{\centering\arraybackslash}m{1.6cm}  >{\centering\arraybackslash}m{1.6cm} >{\centering\arraybackslash}m{2.7cm}  >{\centering\arraybackslash}m{1.3cm}  >{\centering\arraybackslash}m{2.7cm}}
	\hline
	 &  Dice &  AUROC$_{pix}$ & MSE($d,GT$) (var) & SSIM &MSE($r_h, a_h$) (var) \\
	\hline
	StarGAN &   0.710&0.962 &0.0229 (0.128) &0.888 &0.0025 (0.002)\\
	FP-GAN &   0.766&0.975 &0.0160 (0.004) &0.917& 0.0027 (0.003)\\
	VAGAN&  0.442& 0.954 &0.1321 (0.132) &0.869& 0.0036 (0.002)\\
	VAE &  0.288&0.809 &0.0734 (0.071) &0.668& 0.0316 (0.031)\\
	\textbf{DeScarGAN }&  \textbf{0.853}&\textbf{0.988} & \textbf{0.0086(0.002)} & \textbf{0.954}&\textbf{ {0.0018 (0.001)}} \\
	\hline
\end{tabular}
\end{table}

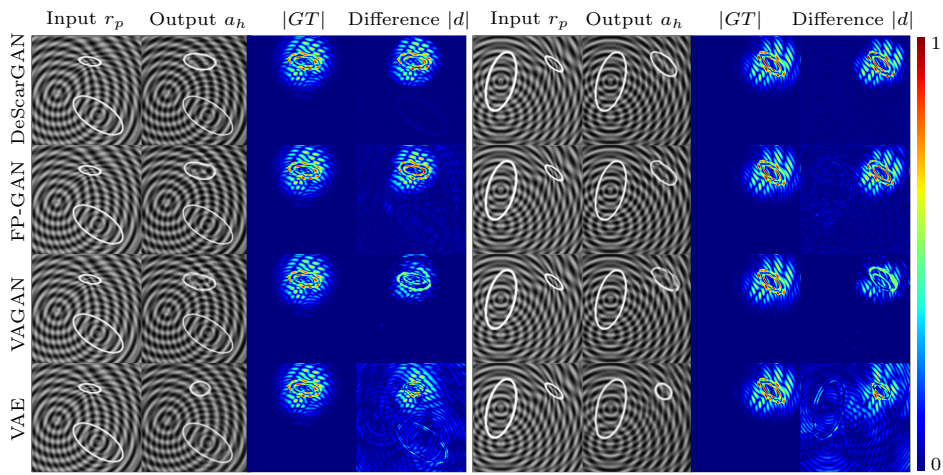
\begin{figure}

	\centering
	\begin{tikzpicture}
	\node[draw=black, inner sep=0pt, thick] at (-0.66, 0) {\includegraphics[scale=0.16]{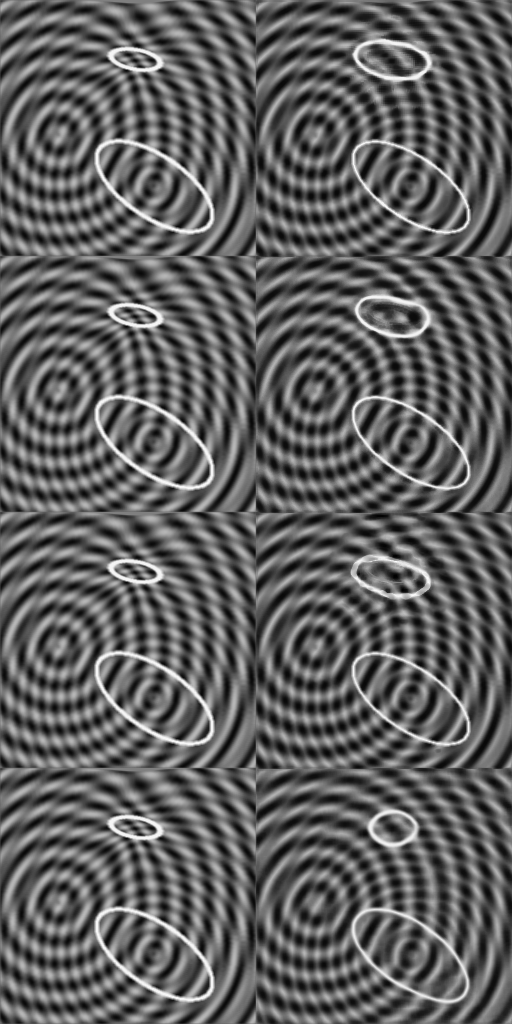}};
	\node[draw=black, inner sep=0pt, thick] at (2.2, 0) {\includegraphics[scale=0.16]{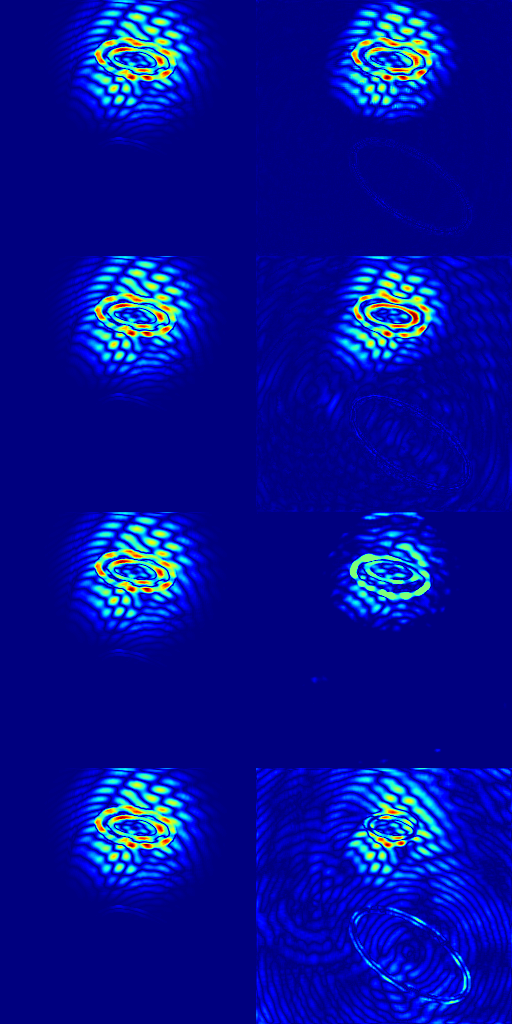}};
	
	\node[] at (-1.4,3.1) {\scriptsize Input $r_p$};
	\node[draw=black, inner sep=0pt, thick] at (5.2, 0) {\includegraphics[scale=0.16]{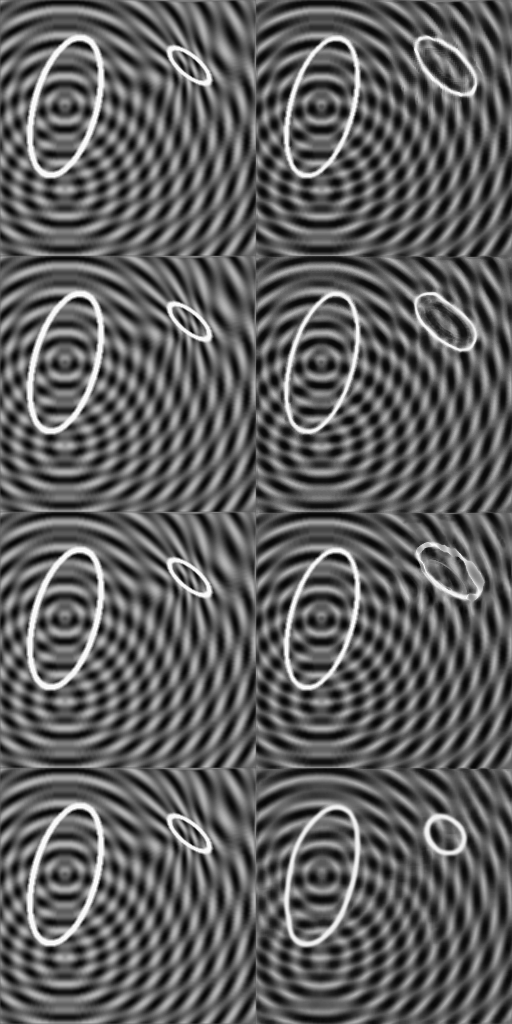}};
	\node[draw=black, inner sep=0pt, thick] at (8.1, 0) {\includegraphics[scale=0.16]{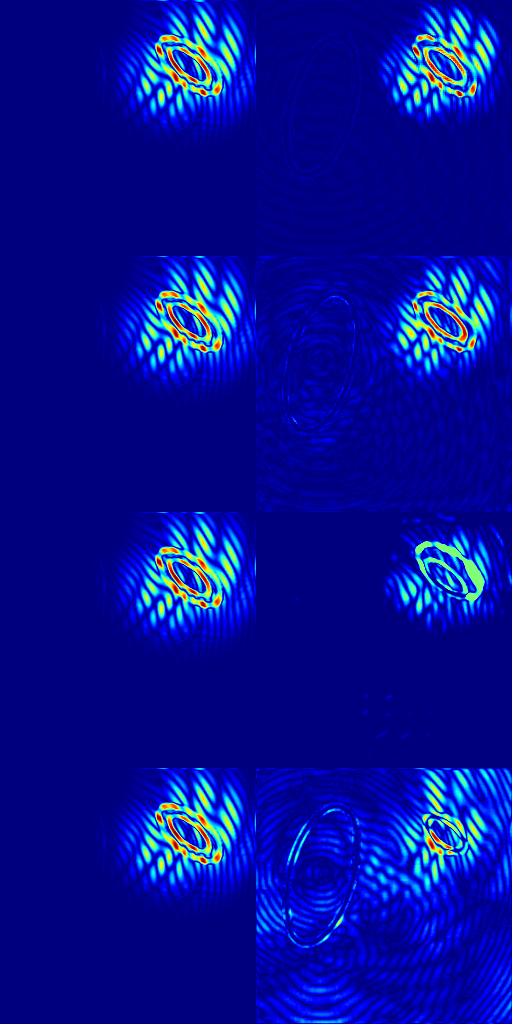}};
	\node[] at (0.1, 3.1) {\scriptsize Output $a_h$};
	
	\node[draw=white, inner sep=0pt, thick] at (9.7, 0) {\includegraphics[scale=0.225]{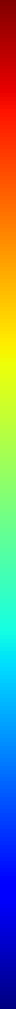}};
	\node[] at (9.9,2.8) {\scriptsize  $1$};
	\node[] at (9.9, -2.8) {\scriptsize  $0$};

	\node[] at (1.5, 3.1) {\scriptsize $|GT|$};
	\node[] at (2.90, 3.1) {\scriptsize Difference $|d|$};

	\node[] at (4.5,3.1) {\scriptsize Input $r_p$};
	\node[] at (5.9, 3.1) {\scriptsize Output $a_h$};

	\node[] at (7.3, 3.1) {\scriptsize $|GT|$};
	\node[] at (8.8, 3.1) {\scriptsize Difference $|d|$};

	\node[rotate=90] at (-2.3,2.2) {\scriptsize DeScarGAN};
	\node[rotate=90] at (-2.3,0.7) {\scriptsize FP-GAN};
	\node[rotate=90] at (-2.3,-0.8) {\scriptsize VAGAN};
	\node[rotate=90] at (-2.3,-2.2) {\scriptsize VAE};
	
	\end{tikzpicture}

	\caption{Visualization of the results of our DeScarGAN, FP-GAN, VAGAN and VAE for two samples of the synthetic dataset. } \label{fig3}	
\end{figure}

In Figure $\ref{fig3}$, exemplary real images ${r_p \in \mathcal{P}}$ of the synthetic dataset with the corresponding artificial images ${a_h \in \mathcal{H}}$ of the different methods are shown.
Our method provides the most accurate difference map $d$. The results of FP-GAN are good as well, but the method fails to generate a proper unshrunken ellipse $e_1$. VAE and VAGAN fail to generate an accurate image of class $h$, resulting in a difference map not close to the ground truth. For visualization, we omit the StarGAN method since it is outperformed by its extension FP-GAN.

\subsection{Chexpert Dataset}
For the Chexpert dataset introduced in \cite{chex}, we used a training set of 14179 images of healthy subjects and 16776 images of subjects that suffer from pleural effusions. The test and validation set each consist of 200 images for each class.

\begin{table}
	\caption{Classification results and  MSE($r_h, a_h$) on the Chexpert dataset. }\label{tab2}
	\centering

	\begin{tabular}{p{2.2cm} >{\centering\arraybackslash}m{2cm}  >{\centering\arraybackslash}m{2cm} >{\centering\arraybackslash}m{2.3cm}  >{\centering\arraybackslash}m{2.8cm} }
		\hline
		 & Accuracy$_{cls}$ & Kappa score &  AUROC$_{image}$ & MSE($r_h, a_h$) (var)  \\
		\hline
		StarGAN &  0.853 & 0.705&0.923 & 0.0534 (0.095)\\
		FP-GAN &  0.875 & 0.750&0.939 &  0.0060 (0.007)\\
		VAGAN& $\times$  &$\times$ &$\times$ & 0.0638 (0.065) \\
		VAE &$\times$ &$\times$ & $\times$&  0.0231 (0.030)\\
		Densenet169 &0.893 & 0.785&0.951& $ \times$\\
		$D_{cls}$  & 0.890 &0.780& 0.949 & $ \times$\\
	
		\textbf{DeScarGAN} &  \textbf{0.898} &\textbf{0.795}&\textbf{0.953} & \textbf{0.0035 (0.003)} \\
		\hline
	\end{tabular}
\end{table}
For classification, we compare DeScarGAN against the classification results of StarGAN, FP-GAN,  Densenet169 \cite{dense} and the classifier $D_{cls}$ without the GAN mechanism. The result for the image-level classification is measured in classification accuracy, the Cohen's kappa score \cite{cohen} and the AUROC score. Further, we measure the MSE between real images $r_h \in\mathcal{H}$ and artificial images $a_h \in \mathcal{H}$. The scores are summarized in Table \ref{tab2}.

\begin{figure}

	\centering
	\begin{tikzpicture}
	\node[draw=black, inner sep=0pt, thick] at (0, 0) {\includegraphics[scale=0.185]{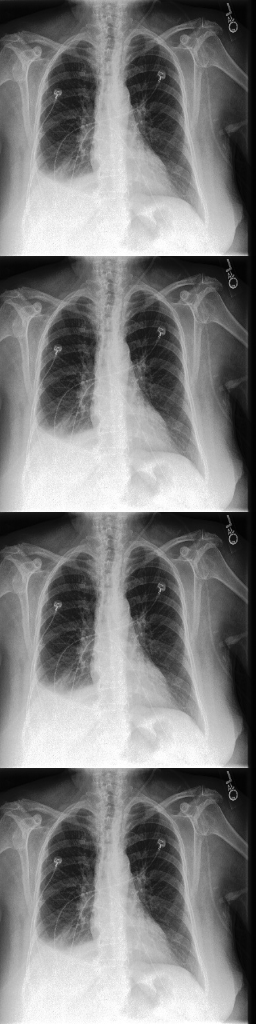}};
	\node[] at (0,3.5)  {\scriptsize Input $r_p$};
	\node[draw=black, inner sep=0pt, thick] at (1.7, 0) {\includegraphics[scale=0.185]{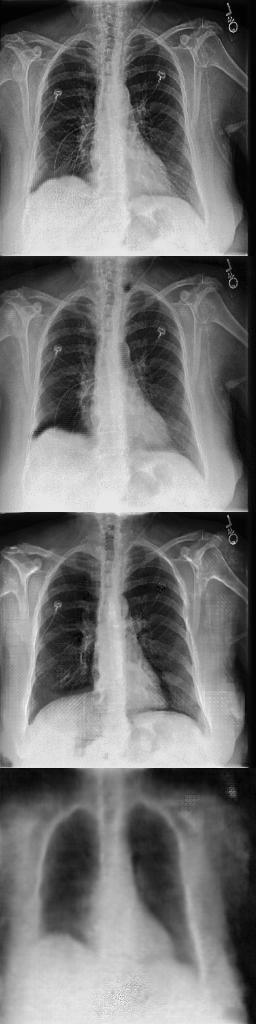}};
	\node[] at (1.7, 3.5)  {\scriptsize Output $a_h$};
	\node[draw=black, inner sep=0pt, thick] at (3.4, 0) {\includegraphics[scale=0.185]{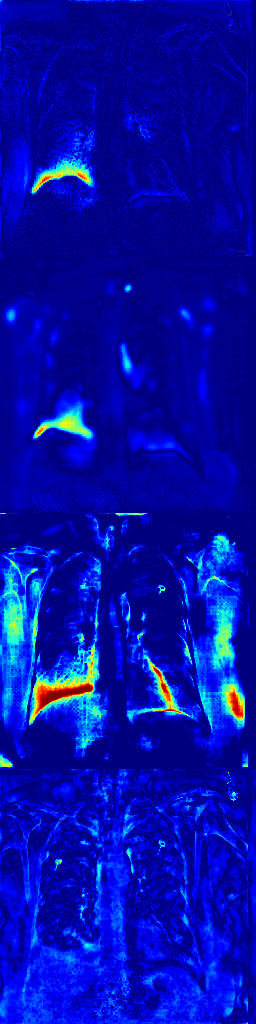}};
	\node[] at (3.4, 3.5) {\scriptsize Difference $|d|$};
	
	\node[draw=black, inner sep=0pt, thick] at (5.6, 0) {\includegraphics[scale=0.185]{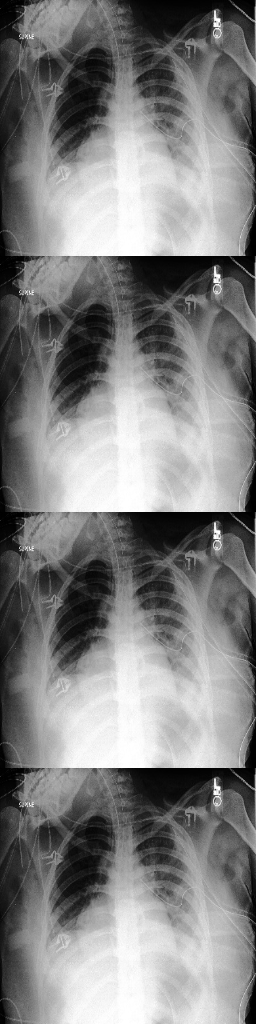}};
	\node[] at (5.6, 3.5) {\scriptsize Input $r_p$};
	\node[draw=black, inner sep=0pt, thick] at (7.3, 0) {\includegraphics[scale=0.185]{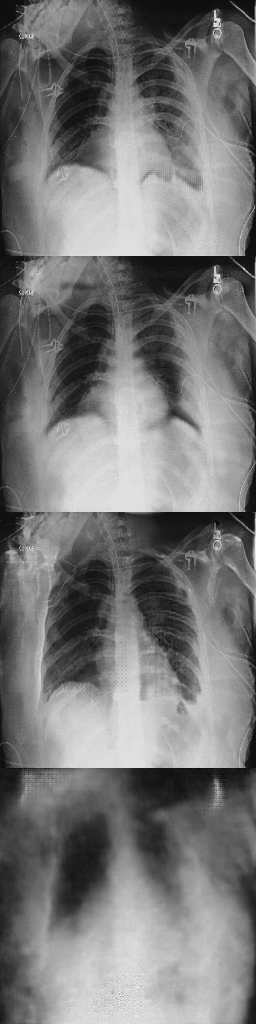}};
	\node[] at (7.3, 3.5) {\scriptsize Output $a_h$};
	\node[draw=black, inner sep=0pt, thick] at (9, 0) {\includegraphics[scale=0.185]{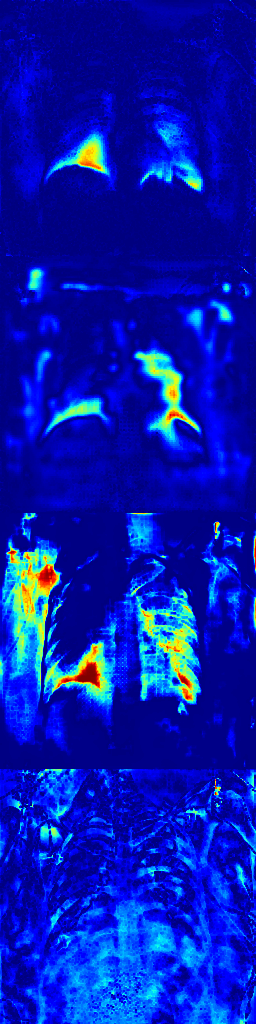}};
	\node[] at (9, 3.5){\scriptsize Difference $|d|$};
	
	\node[draw=white, inner sep=0pt, thick] at (10.2, 0) {\includegraphics[scale=0.262]{cbar3}};
	\node[] at (10.5, 3.15) {\scriptsize  $1$};
	\node[] at (10.5, -3.15) {\scriptsize  $0$};
	1
    \node[rotate=90] at (-1.1,2.5) {\scriptsize DeScarGAN};
    \node[rotate=90] at (-1.1,0.85) {\scriptsize FP-GAN};
    \node[rotate=90] at (-1.1,-0.8) {\scriptsize VAGAN};
    \node[rotate=90] at (-1.1,-2.5) {\scriptsize VAE};
    
	\end{tikzpicture}

\caption{Comparison of our DeScarGAN against FP-GAN, VAGAN and VAE for two samples of the Chexpert dataset.} \label{fig4}	
\end{figure}

DeScarGAN achieves better classification results than the pure classification networks $D_{cls}$ and Densenet169, indicating that the GAN mechanism supports the classification network. The results of the different methods are visualized in Figure~\ref{fig4}. We observe that the VAE fails to detect pleural effusions. Although FP-GAN  detects similar regions as our method, the generated maps appear blurry and mark regions outside the thorax. The additive map of VAGAN also outlines parts of the arms and upper chest as abnormal. Our method generates the most detailed difference map, not highlighting any regions outside the pleural space.

\section{Conclusion}
We proposed DeScarGAN, a method to generate disease-specific, detailed maps that show pathological changes of existing anatomical structures. The novelty of our method is the introduction of a new architecture with skip connections, a splitting of the networks into weight-sharing subnetworks and an identity loss as identity-preserving mechanisms. This setup enables the detection of deformations of existing anatomical structures, e.g., atrophy or changes in the pleural space due to pleural effusions. \\
When comparing our DeScarGAN against state-of-the-art anomaly detection algorithms, we outperform  FP-GAN, VAE, StarGAN and VAGAN on a synthetic dataset. Although FP-GAN provides good results by generating additive maps, DeScarGAN generates a complete new image and provides more precise maps that reliably outline the regions of pathological changes.\\
When applying our model on the Chexpert lung X-ray dataset with pleural effusions, our classification scores are better than state-of-the-art classification networks. The generated maps detect anomalies in a detailed manner and lead the attention to the relevant parts of the anatomy.
This approach has the potential to bridge the gap between the knowledge about the presence of a disease and setting the focus of a longitudinal study observing the region of interest.

\subsubsection{Acknowledgement.}
This work was supported by Novartis FreeNovation.

\bibliographystyle{splncs04}
\bibliography{bib_julia}

\end{document}

\title{Supplementary Material to DeScarGAN}

\author{
Julia Wolleb \and
Robin Sandk\"uhler \and
Philippe C. Cattin}

\institute{}
\authorrunning{J. Wolleb et al.}

\maketitle              

\section{ Generation of the Synthetic Dataset}

\begin{table}
	\caption{Equations used for the generation of the synthetic dataset. The images are of size $256\times256$, and the image coordinates run from $0$ to $255$. The ellipses have a rotation angle $\phi$ and a contour thickness $g$. If the two ellipses intersect, the image is excluded from the dataset. The images of the healthy group $\mathcal{H}$ show the two ellipses $e_1$ and $e_2$, the background is given by the circles $c_1$ and $c_2$. The images of the diseased group $\mathcal{P}$ are generated by deforming a healthy image with the given deformation field.
	}

	\centering
	\begin{tabular} {p{2cm} >{\arraybackslash}m{5.8cm}  >{\arraybackslash}m{5.2cm}}
		\hline
	name	& formula &  parameters\\
		\hline
		ellipse $e_1$ & $(a \cos(\theta)+m_1, b \sin(\theta)+m_2)=1$  & ${m_1,m_2 \thicksim U([50,200])} $, ${\forall \theta \in [0,2\pi)}$, ${a\thicksim \mathcal{N}(40,1)}$, ${b\thicksim \mathcal{N}(20,1)}$, ${\phi\thicksim U([0,2\pi])}$, ${g\thicksim \mathcal{N}(5,0.7)}$\\
			\hline
	ellipse $e_2$ & $(a \cos(\theta)+m_1, b \sin(\theta)+m_2)=1$  & ${m_1,m_2  \thicksim U([50,200])}$, ${\forall \theta \in [0,2\pi)}$, ${a\thicksim \mathcal{N}(70,1)}$, ${b\thicksim \mathcal{N}(35,1)}$, ${\phi\thicksim U([0,2\pi])}$, ${g\thicksim \mathcal{N}(5,0.7)}$\\
		\hline
		circle $c_1$& $(t\cos\theta+m1, t\sin\theta+m_2)=\sin(tf)h$& ${\forall t \in [0,255]}$, ${\forall \theta \in [0,2\pi)}$,  $h{\thicksim U([0.2,0.4])}$, ${f\thicksim U([0.2,0.35])}$, ${m_1,m_2\thicksim U([50,200])}$\\
			\hline
			circle $c_2$& $(t\cos\theta+m1, t\sin\theta+m_2)=\sin(tf)h$& ${\forall t \in [0,255]}$, ${\forall \theta \in [0,2\pi)}$, ${h\thicksim U([0.2,0.4])}$, ${f\thicksim U([0.35,0.5])}$, $m_1$ and $m_2$ of $e_2$\\
				\hline
		
		deformation field&$\nabla \Bigg(\cfrac{ 1} {(2  \pi * 0.19 ^2)} \;  e^{-\big(\frac{(x - m_1)^2}{(2 * 0.19^2)} +\frac{(y - m_2)^2}{ (2 * 0.19^2)}\big)}\Bigg)$ & $m_1$ and $m_2$ of $e_1$\\
		\hline
	\end{tabular}
\end{table}

\begin{figure}

	\centering
	\begin{tikzpicture}
	\node[draw=white, inner sep=0.1pt, thick] at (1.5,0) {\includegraphics[scale=0.29]{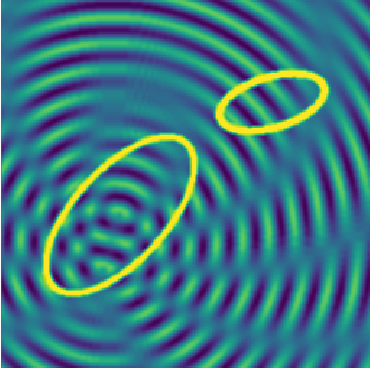}};
	\node[] at (1.5, 1.8) {\small(a)};

	\node[draw=white, inner sep=0.1pt, thick] at (5.6,0){\includegraphics[scale=0.14]{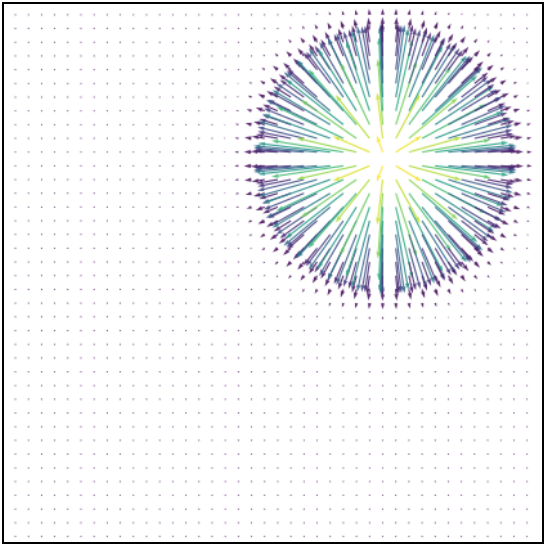}};

	\node[] at (5.6, 1.8) {\small(b)};
	\node[draw=white, inner sep=0.1pt, thick] at (9.6, 0)  {\includegraphics[scale=0.29]{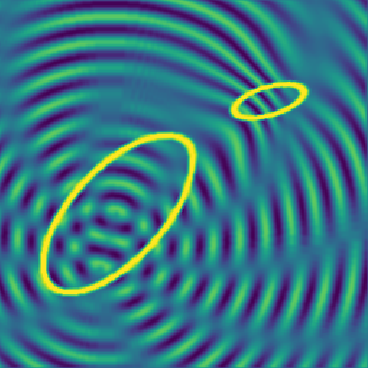}};
	\node[] at (9.6, 1.8) {\small(c)};
	
		\node[draw=white, inner sep=0.1pt, thick] at (1.5,-3) {\includegraphics[scale=0.3]{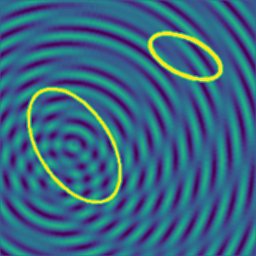}};
		\node[draw=white, inner sep=0.01pt, thick] at (5.6,-3){\includegraphics[scale=0.14]{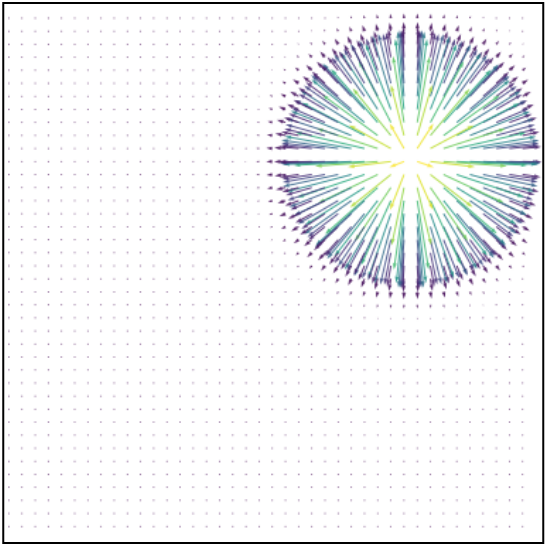}};
		\node[draw=white, inner sep=0.1pt, thick] at (9.6, -3)  {\includegraphics[scale=0.3]{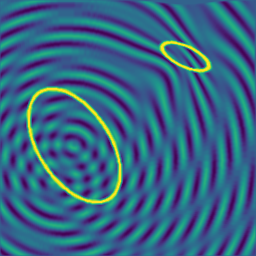}};
		
	\node[draw=white, inner sep=0.1pt, thick] at (1.5,-6) {\includegraphics[scale=0.3]{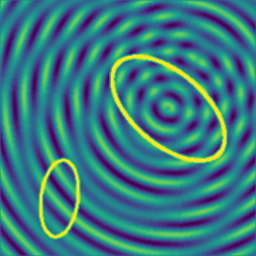}};
	\node[draw=white, inner sep=0.1pt, thick] at (5.6,-6){\includegraphics[scale=0.14]{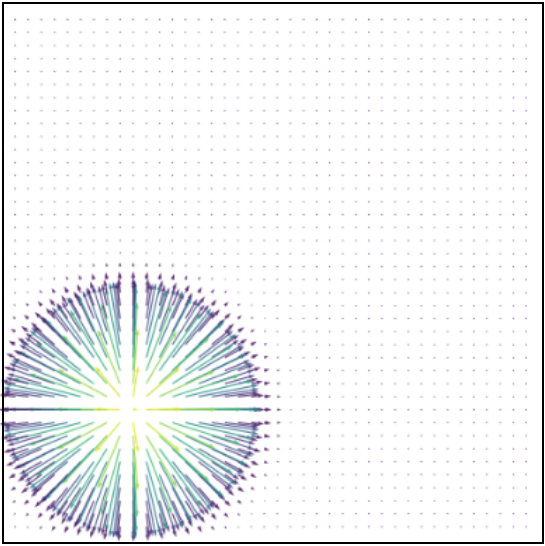}};
	\node[draw=white, inner sep=0.1pt, thick] at (9.6, -6)  {\includegraphics[scale=0.3]{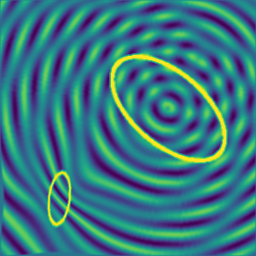}};	
	
	\end{tikzpicture}

	\caption{Image (a) shows an exemplary  image of the set $\mathcal{H}$. Image (b) shows the deformation field used to transform image (a) into image (c) of the  diseased group $\mathcal{P}$. For the dataset, either image (a) or image (c) is chosen, such that the sets $\mathcal{H}$ and $\mathcal{P}$ remain unpaired.}
	\label{fig2}
\end{figure}